\documentclass[aps,prb,showpacs,reprint]{revtex4-1}

\usepackage{graphicx}
\usepackage{amsmath}
\usepackage{bbm}

\begin{document}
\title
{Resonant Landau-Zener transitions in helical magnetic fields}
\author{P. W\'ojcik}
\email[Electronic address: ]{pawelwojcik@fis.agh.edu.pl}
\author{J. Adamowski}
\author{M. Wo{\l}oszyn}
\author{B. J. Spisak}
\affiliation{AGH University of Science and Technology, Faculty of
Physics and Applied Computer Science, al. Mickiewicza 30,
Krak\'ow, Poland}

\begin{abstract}
The spin-dependent electron transport has been studied in magnetic
semiconductor waveguides (nanowires) in the helical magnetic
field.  We have shown that -- apart from the known conductance dip
located at the magnetic field equal to the helical-field amplitude $B_h$
-- the additional conductance dips (with zero conductance) appear at
magnetic field different from $B_h$. This effect occuring in the
non-adiabatic regime is explained as resulting from the resonant
Landau-Zener transitions between the spin-splitted subbands. 

\end{abstract}

\pacs{xxx}

\maketitle
The experimental realization of an effective spin-transistor remains a
challenge facing spintronics since the pioneering concept proposed by
Datta and Das.\cite{Datta1990} According to the original
idea the operation of the spin transistor is based on
the gate-controlled spin-orbit interaction (SOI) of Rashba
form.\cite{Rashba1984} The current of spin polarized electrons is
injected from the ferromagnetic source into the conduction channel
formed in a two-dimensional electron gas (2DEG) and is ballistically
transported to the ferromagnetic drain.  
The state of the transistor depends on the electron spin
orientation modulated via the Rashba SOI by the voltage applied to the
gate attached close to the channel.
The operation of spin transistor has been studied in many theoretical
papers.~\cite{Wang2004,Zhang2005,Kunihashi2012,Wojcik2014,Wojcik}
However, the experiments\cite{Koo2009,Yoh2012} indicate that the signals
obtained in the up-to-date realized spin transistors based on the SOI
are rather low, which results from the low efficiency of the spin
injection from the ferromagnet into the semiconductor\cite{Schmidt2000}
and the spin relaxation. The SOI  causes that the scattering processes
affect the spin states of electrons,  e.g., by the Elliott-Yafet or
Dyakonov-Perel mechanism.\cite{Perel1971} 
Although the spin relaxation is proposed to be suppressed by equating
the Rashba and Dresselhaus term,\cite{Schliemann2003, Koralek2009} the
concept of the non-ballistic spin transistor proposed by Schliemann et
al. in Ref.~\onlinecite{Schliemann2003} is still waiting for the
experimental realizations.

An alternative spin-transistor design, with the spin relaxation length
as long as 50~$\mu$m, has been recently described by {\v Z}uti{\'c} and
Lee\cite{Zutic2012} and experimentally demonstrated by Betthausen et
al. in Ref.~\onlinecite{Betthausen2012}. In this approach, the spin
control is realized by combining the homogeneous and helical magnetic
fields. The latter is generated by ferromagnetic stripes
located above the conduction channel. The spin state of electrons
flowing through the channel is protected against a possible decay by
keeping the transport in the adiabatic regime.\cite{Frustaglia2001} The
transistor action is driven by the diabatic Landau-Zener
transitions\cite{Landau,Zener} induced by the appropriate tuning of the
homogeneous magnetic field.  For these suitably chosen conditions the
backscattering of spin polarized electrons appears, which gives raise to
the large increase of the resistance, i.e., the transistor goes over
into the 'off' state. In contrast to the SOI-based spin transistor, the
proposed design is robust against the scattering
processes.\cite{Saarikoski2014}

Motivated by the experiment,~\cite{Betthausen2012} we have performed
the computer simulations of the spin-dependent transport
in the helical magnetic field for the nanostructure similar to that of
Ref.~\onlinecite{Betthausen2012}. The calculations of the conductance as
a function of the Fermi energy reveals additional conductance dips,
which cannot be understood in terms of the ordinary Landau-Zener
theory. This effect, not reported in the previous
studies,\cite{Betthausen2012,Saarikoski2014} is
explained as resulting from the resonant inter-subband transition
between the spatially modulated spin-dependent subbands. \\
\begin{figure}[ht]
\begin{center}
\includegraphics[scale=0.35, angle=0]{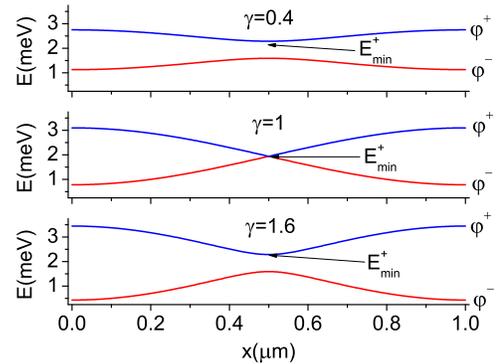}
\caption{(Color online)
Electron potential energy profiles for different values of $\gamma=
B_{ext}/B_h$. $E$ is the minimum energy of the electron in
subband $\varphi^{\pm}$ ($+$ and $-$ denotes the state with spin
parallel and antiparallel with respect to the magnetic field),
coordinate $x$ is measured along the electron
flow.
}
\label{fig1}
\end{center}
\end{figure}

We consider a two-dimensional  waveguide (nanowire)
in the $x-y$ plane made of (Cd,Mn)Te in the presence of the magnetic
field $\mathbf{B}(\mathbf{r})$, which is the superposition of the
homogeneous magnetic field applied along the $z$-axis
$\mathbf{B}_{ext}=(0,0,-B_{ext})$ and the helical magnetic
field $\mathbf{B}_h(\mathbf{r})$ taken on in the
form\cite{Saarikoski2014}
\begin{equation}
\mathbf{B}_h(\mathbf{r})=B_h \left ( \sin {\frac{2 \pi x}{a}}, \cos
{\frac{2 \pi x}{a}}, 0 \right ) \;,
\end{equation}
where $a$ is the period of the magnetic field modulation.
The Hamiltonian of the electron is given by
\begin{equation}
\hat{H}=\frac{1}{2m_{e\!f\!f}} \left [
\hat{\mathbf{p}}+e\mathbf{A}(\mathbf{r})
\right ] ^2 + \frac{1}{2}g_{e\!f\!f}\mu_B \mathbf{B}(\mathbf{r}) \cdot
\pmb{\sigma} \;,
\end{equation}
where $m_{e\!f\!f}$ is the conduction-band mass, $\mathbf{A}(\mathbf{r})
= (\mathbf{B}\times \mathbf{r})/2$ is the vector potential,
$g_{e\!f\!f}$ is the effective $g$-factor, $\mu_B$ is the Bohr
magneton, and $\pmb{\sigma}$ is the vector of Pauli matrices.
In the presence of the magnetic field, the $s$-$d$ exchange
interaction between the conduction band electrons and Mn ions in
(Cd,Mn)Te
leads to the giant Zeeman splitting of the conduction bands with the
effective $g$-factor ranging from 200 to 500.\cite{Furdyna1988}
\begin{figure}[!ht]
\begin{center}
\includegraphics[scale=0.2, angle=0]{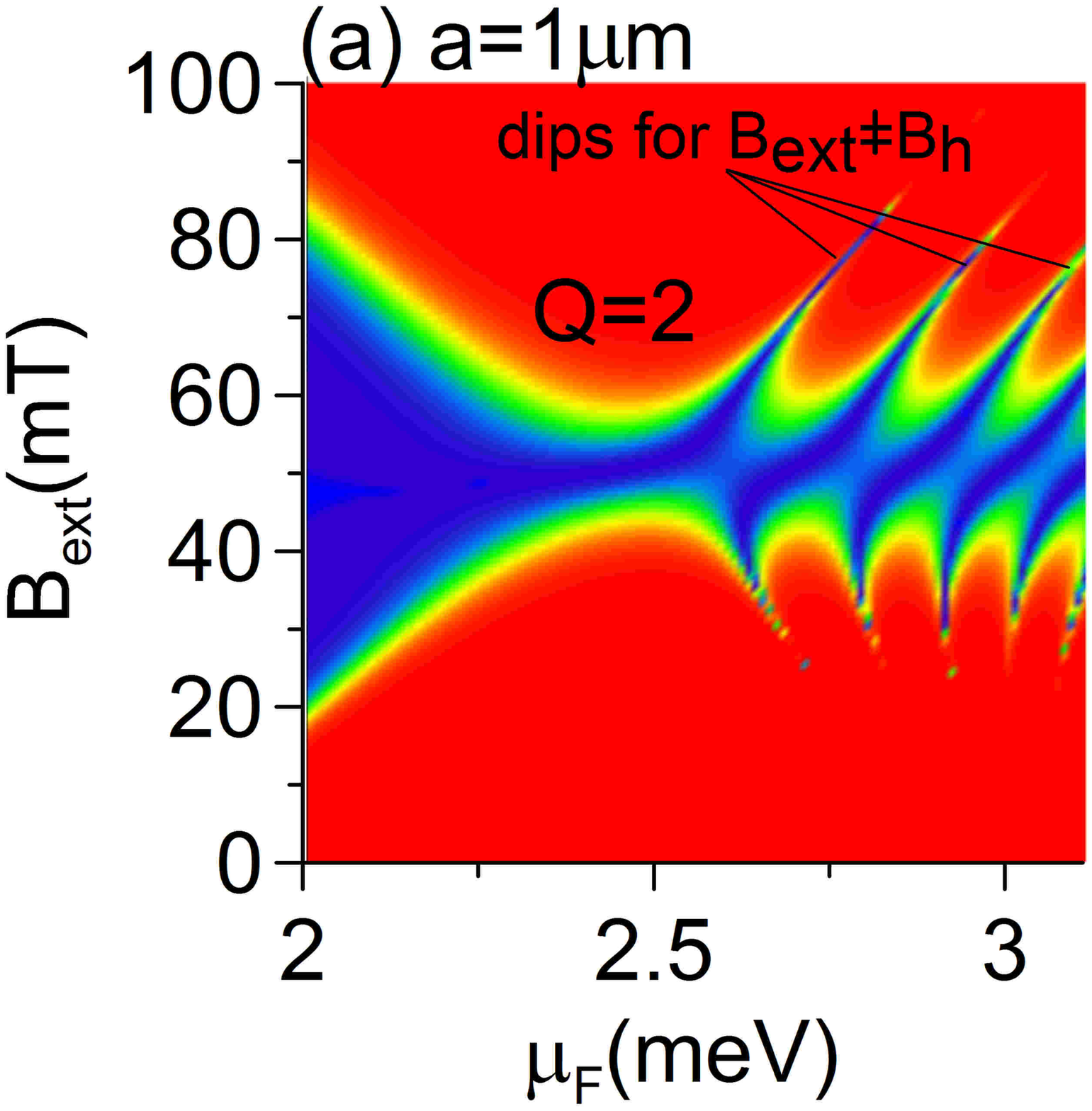}
\includegraphics[scale=0.2, angle=0]{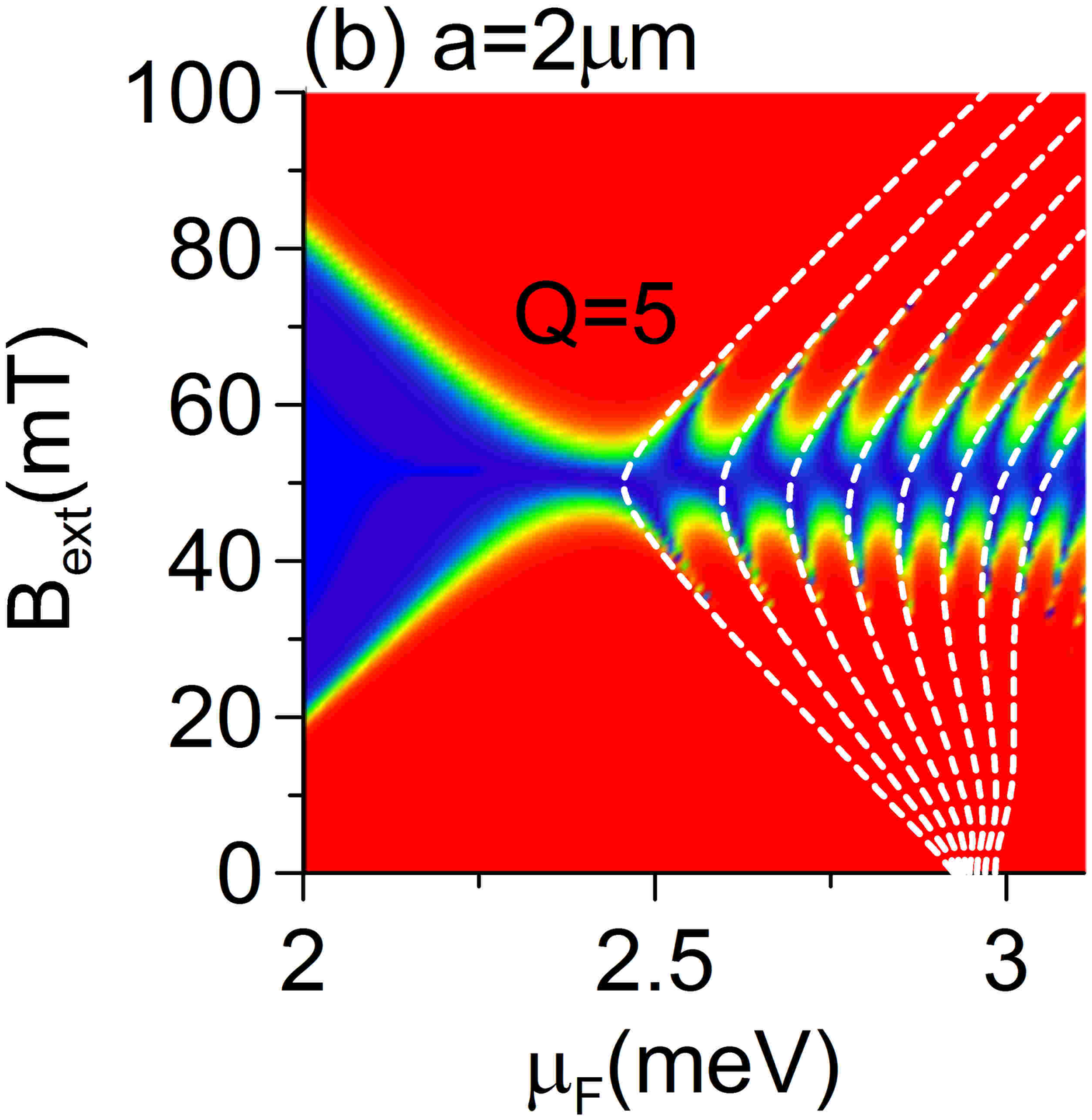}
\includegraphics[scale=0.26, angle=0]{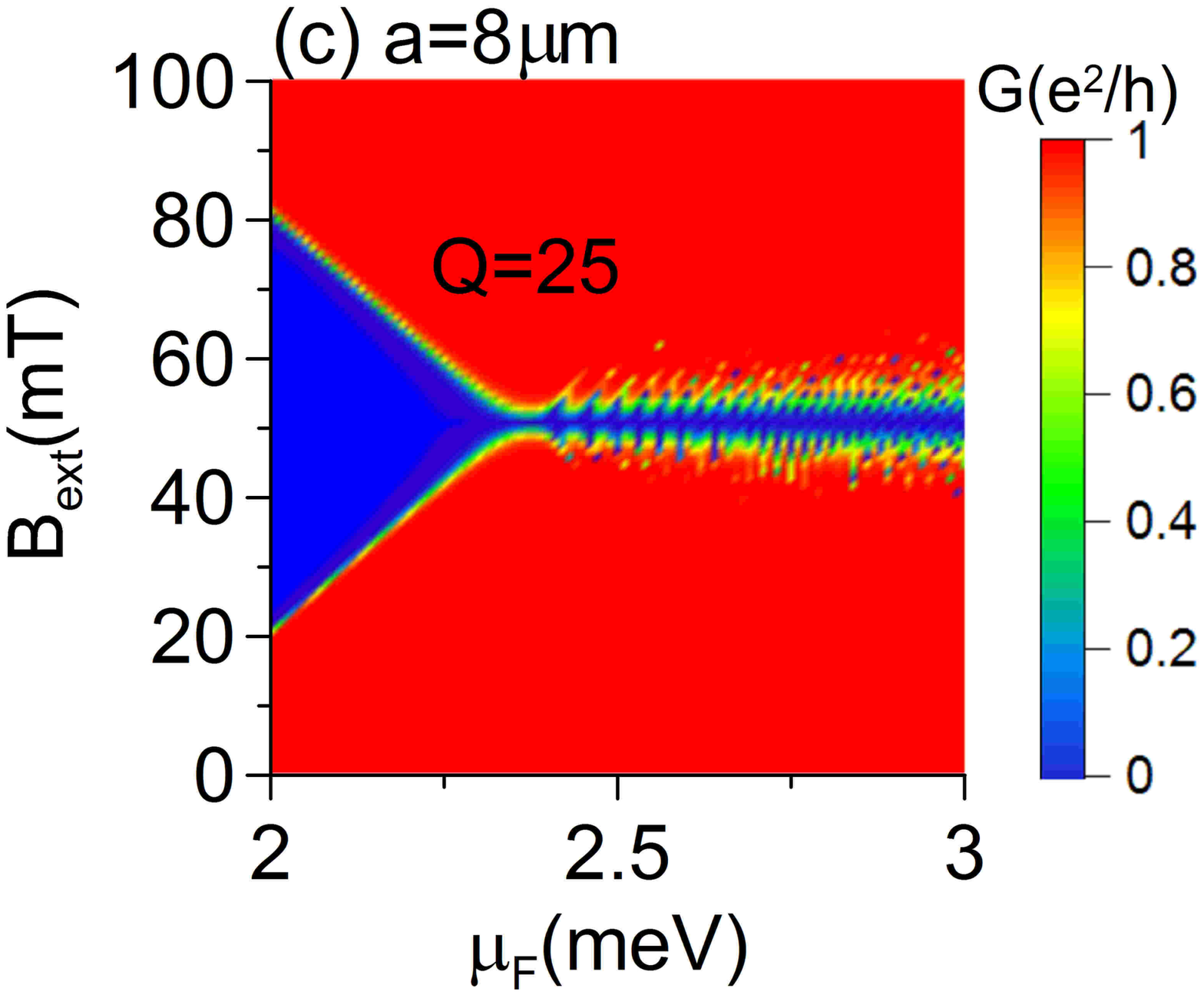}
\caption{(Color online) Conductance $G$ as a function of homogeneous
magnetic field $B_{ext}$ and Fermi energy $\mu_F$.
The values of the period of magnetic field modulation (a) $a=1 \mu$m,
(b) $a=2 \mu$m, and (c) $a=8\mu$m correspond to
$Q=\omega_L/\omega_{mod}=2,5$ and $25$, respectively.
The white dashed lines correspond to the energies of the quasi-bound
states in the effective quantum well seen by the electrons in
subband $\varphi ^{+}$.}
\label{fig2}
\end{center}
\end{figure}

In the superposition of the magnetic fields, the subbands corresponding
to the electron with the parallel $\varphi^{+}$ and antiparallel
$\varphi^{-}$ spin orientation (with respect to the magnetic field)
depend on the postion $x$. The spatially-dependent subband energy
minima $E^{\pm}$ are given by [cf. Fig.~\ref{fig1}]
\begin{equation}
E^{\pm} (x) =\pm \frac{1}{2} g_{e\!f\!f} \mu_B B_{h} \sqrt{1+\gamma ^2
+ 2 \gamma \cos {(2\pi x/a)}} \;,
\end{equation}
where $\gamma=B_{ext}/B_h$.  Parameter $\gamma$ that can be tuned by
changing the homogeneous magnetic field $B_{ext}$ determines the spatial
modulation of the spin-splitted subbands as depicted in Fig.~\ref{fig1}.

We have performed the numerical calculations of the conductance by the
tight-binding method on the square lattice with $\Delta x = \Delta
y=1$~nm using the Kwant package\cite{kwant}.
In the calculations, the following values of the parameters have been
used: the effective mass of electron in CdTe, $m_{e\!f\!f}=0.1 m_e$
where $m_e$ is the free electron mass, and $g_{eff}=200$.
We adopt the hard-wall boundary conditions in the $y$ direction
assuming the width of the conduction channel $W=30$~nm. The value of the
helical magnetic field amplitude $B_h$ has been taken on the basis of
the experimental report\cite{Betthausen2012} and is equal to
$B_h=50$~mT. 

The previous studies\cite{Betthausen2012,Saarikoski2014} of the spin
control in the helical magnetic field have been performed in
the adiabatic regime, in which the spin orientation of electrons flowing
through the nanostructure follows the spatial modulation of the magnetic
field. In the considered nanostructure, the
adiabaticity\cite{Frustaglia2001} can be defined by the parameter
$Q=\omega_L/\omega _{mod}$, where $\omega_L=g_{eff}\mu_B B/\hbar$
is the frequency of the spin Larmor precession and $\omega _{mod}=2\pi
v_F/a$ is the frequency of the magnetic field modulation measured in the
electron rest frame, where $v_F$ is the Fermi velocity. In the adiabatic
regime, $Q\gg 1$.

\begin{figure}[ht]
\begin{center}
\includegraphics[scale=0.3, angle=0]{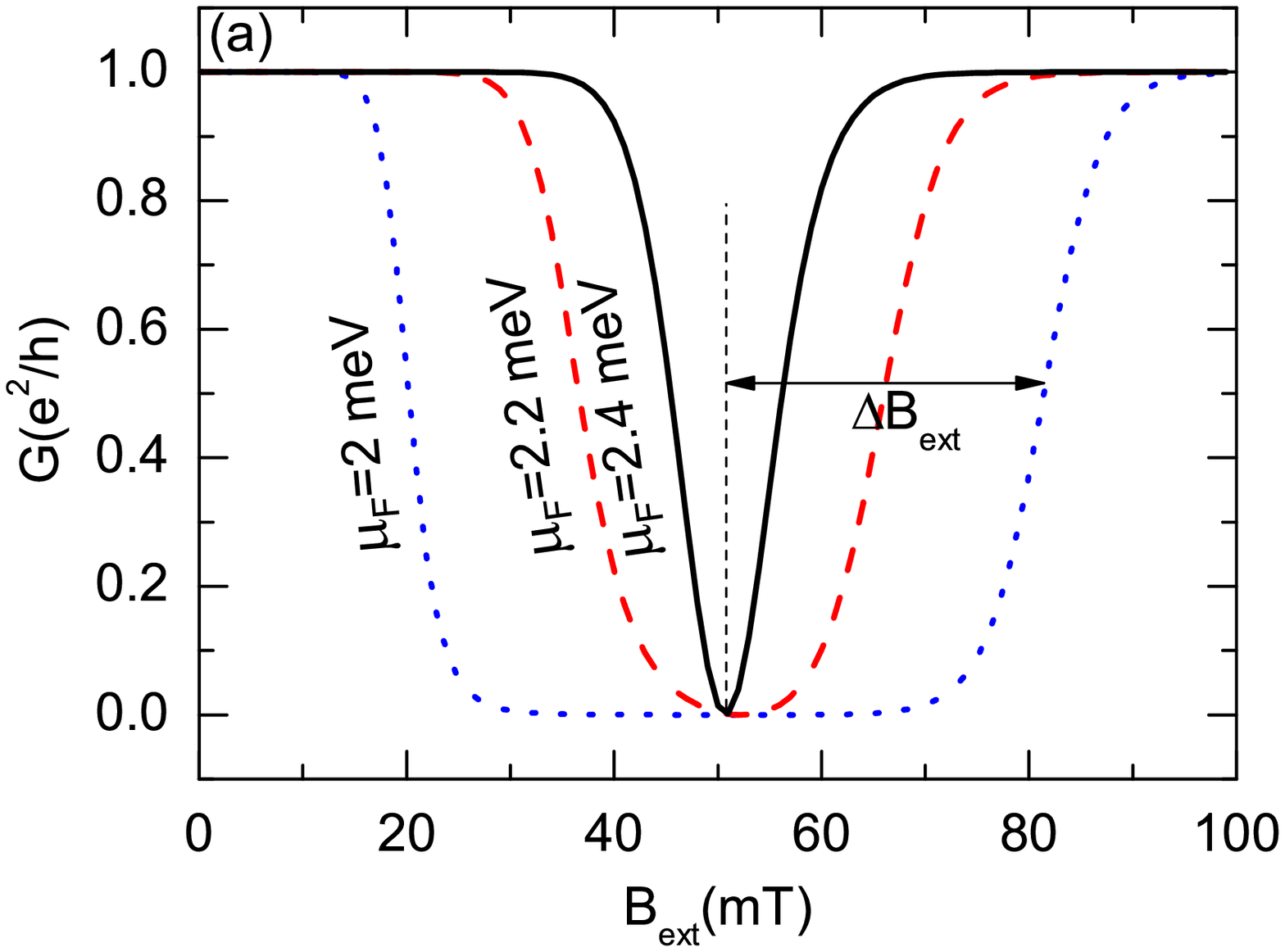}
\includegraphics[scale=0.3, angle=0]{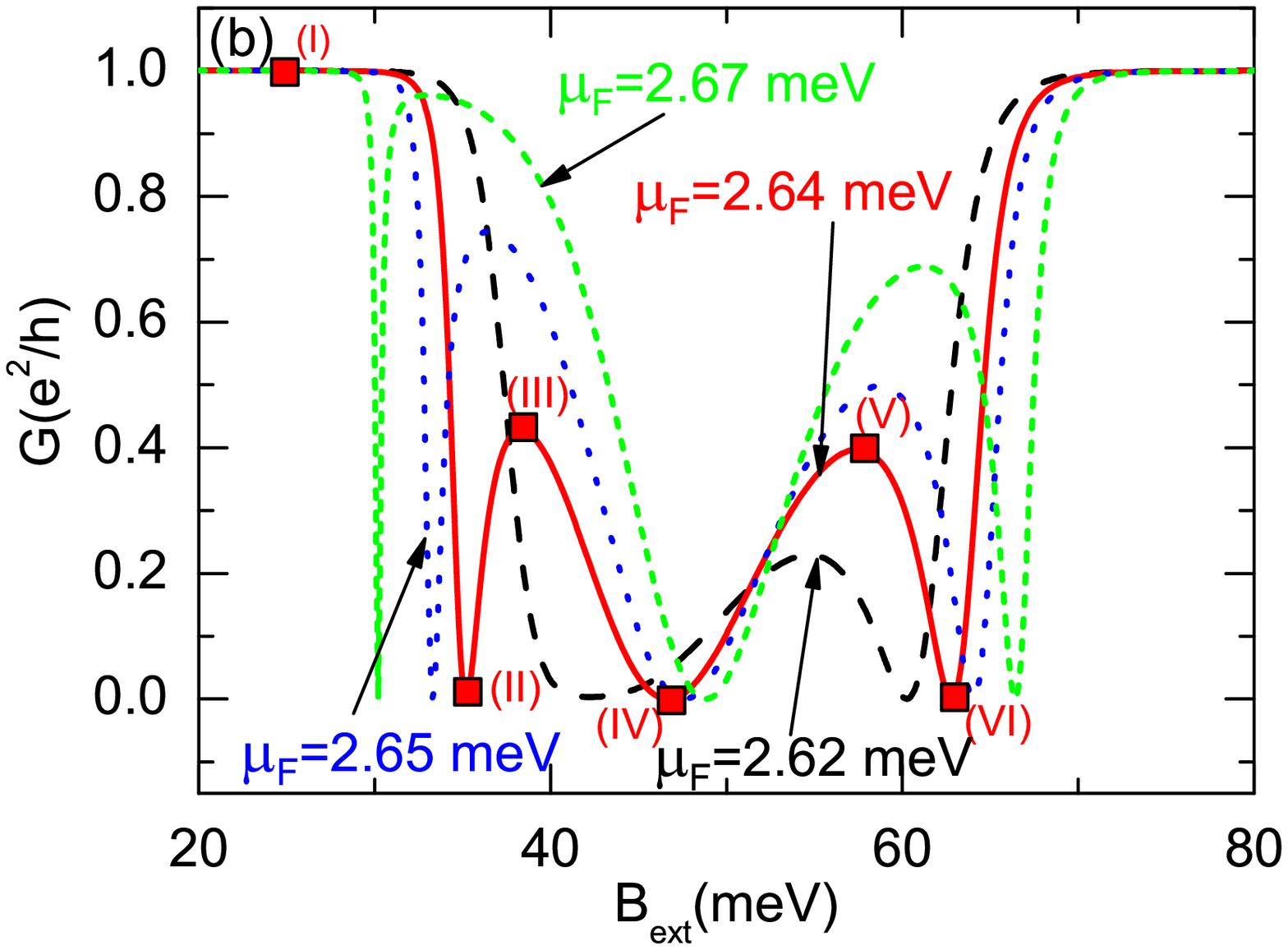}
\caption{(Color online) Cross-sections of the conductance maps for
Fermi energies (a) $\mu_F < E_{min}^{+}$  and (b) $\mu_F >E_{min}^{+}$.
In panel (b), the additional conductance dips for $B_{ext}\ne
B_h$ appear [these additional dips are marked as points (II) and (VI)].
Results for $a=1\mu$m.}
\label{fig3}
\end{center}
\end{figure}

In this paper, we study the spin transport in the regime, in which the
adiabaticity condition is weakened.  In order to achieve this, we 
change the period $a$ of the helical-field modulation.
Figure~\ref{fig2} presents the conductance as a function of the
homogeneous magnetic field $B_{ext}$ and  the Fermi energy
$\mu _F$ for the different values of $a$ corresponding to different $Q$.
In the calculations, we have considered only the two
lowest-energy spin-splitted subbands.  The Fermi energy is varied within
the range, which ensures that the electrons are injected into the
conduction channel from the lowest-energy subband. The conductance as a
function of the magnetic field and Fermi energy exhibits the wide
minimum centered around the value of helical magnetic field amplitude
$B_h$ [Figs.~\ref{fig2} and \ref{fig3}].
The conductance minimum in the range of the magnetic
field $(B_h - \Delta B_{ext}, B_h + \Delta B_{ext})$ can be easily
understood in terms of the Landau-Zener transitions between the
spatially modulated Zeeman splitted energy subbands.
The value of $\Delta B_{ext}$ depends on the position of Fermi energy
with respect to the spatially dependent energy subbands.
The mechanism leading to the central wide dip has been explained in
details in Refs.~\onlinecite{Betthausen2012,Saarikoski2014}.

\begin{figure}[ht]
\begin{center}
\includegraphics[scale=0.3, angle=0]{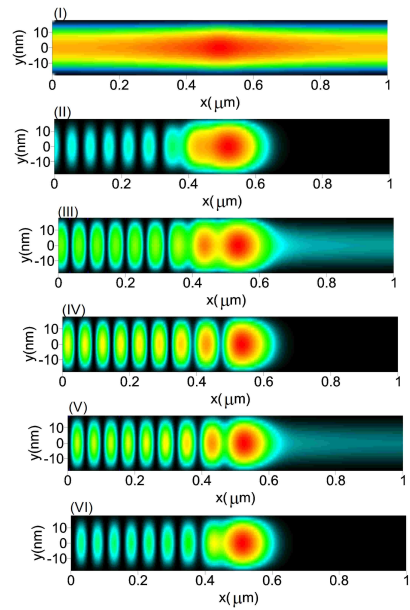}
\caption{(Color online) Electron density in the nanostructure calculated
for magnetic fields marked by (I)-(VI) in Fig.~\ref{fig3}(b).
}
\label{fig4}
\end{center}
\end{figure}

\begin{figure}[ht]
\begin{center}
\includegraphics[scale=0.4, angle=0]{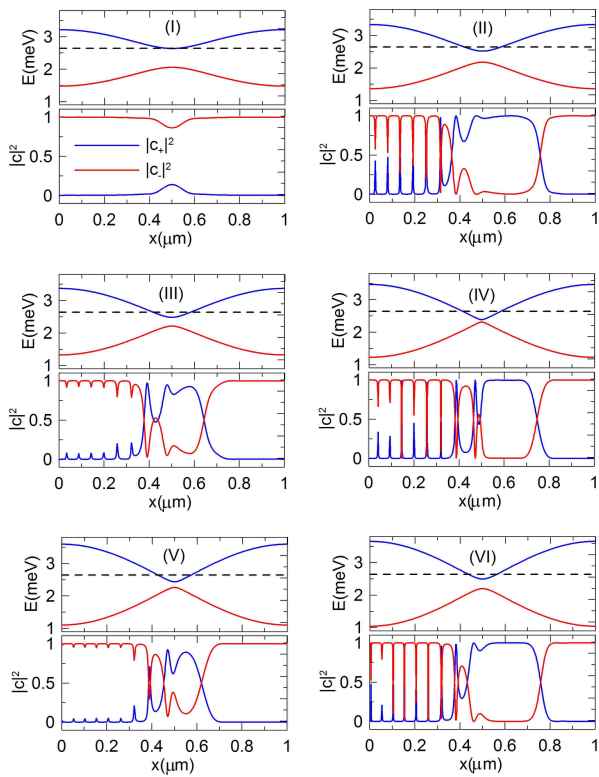}
\caption{(Color online) Spatially varying minima $E^{\pm}(x)$ of the
conduction subband energy
(upper panels) and the contribution of each subband to the wave function
$|c_1|^2$ (red line) and $|c_2|^2$ (blue line) calculated for the
magnetic fields (I)-(VI) marked in Fig.~\ref{fig3}(b).
}
\label{fig5}
\end{center}
\end{figure}

In our study we have found that if the Fermi energy exceeds 
$E_{min}^{+}$ defined as the minima of the band $\varphi ^{+}$ (see
Fig.~\ref{fig1}), the additional conductance dips
appear for $B_{ext}\ne B_h$  [see Fig.~\ref{fig2}(a,b) and
Fig.~\ref{fig3}(b)]. This effect has not been reported in the previous
studies\cite{Betthausen2012, Saarikoski2014} and cannot be explained
as resulting from the ordinary Landau-Zener transitions, for which the
conductance is suppressed only for $B_{ext} \simeq B_h$.  The analysis
of the conductance maps in Fig.~\ref{fig2}(a,b) allows us to distinguish
two characteristic features of this effect: (i) the period of the
occurrence of the additional dips (measured as a function of the Fermi
energy) increases with the increasing period $a$ of the magnetic field
modulation [cf. Fig.~\ref{fig2}(a) and Fig.~\ref{fig2}(b)], (ii) the
effect disappears in the adiabatic regime (we see that for $a=8$~$\mu$m
($Q\approx25$) the additional conductance dips are strongly suppressed
[cf. Fig.~\ref{fig2}(c)]).
In order to find the physical interpretation of these additional minima
of conductance, we have analyzed  six characteristic values of the
magnetic field marked by labels (I)-(VI) on the red curve in
Fig.~\ref{fig3}(b). The chosen values of the magnetic field correspond
to the subsequent maxima and minima of the conductance. Figure
\ref{fig4} displays the electron density in the
nanostructure calculated for points (I)-(VI).
For the case (I) we observe the transmission of electrons through
the waveguide with the slight increase of the charge density in the
vicinity of the point $x=a/2$. At this point the electron transmission
is completely blocked in the cases (II), (IV), and (VI). The partial
transmission with the non-zero electron density for $x > a/2$ has been
obtained in the cases (III) and (V). Figure \ref{fig4} shows that the
three conductance dips correspond to the backscattering of electrons in
the vicinity of $x=a/2$. We note that the conductance dips for $B_{ext}
\ne B_h$ appear only in the non-adiabatic regime for the short period
$a$, which corresponds to $Q < 5$. In this regime, the modulated
magnetic field, acting on the electron in its reference frame, changes
so fast that the electron spin cannot adapt to changes of the magnetic
field. This means that the electron, initially injected from the lowest
subband $\varphi ^{-}$, does not remain in this subband when flowing
through the nanostructure - the electron quantum state can be described
by the linear combination of the eigenstates with the spin orientation
parallel and antiparallel to the magnetic field, i.e.,
by $\Psi(x,y)=c_+(x)\varphi^{+}(y)+c_-(x)\varphi^{-}(y)$, where
$c_+(x)$ and $c_-(x)$ are the space-dependent coefficients.
The contribution of each subband  $|c_{\pm}(x)|^2$
is presented in Fig.~\ref{fig5}. We see that for the magnetic fields
(II), (IV), (VI), i.e., for the conductance dips, the contribution to
the wave function in the vicinity of $x = a/2$ originates only from the
subband $\varphi^{+}$.
Since the electrons are injected into the nanowire in state $\varphi
^{-}$, related to the lowest-energy subband, one can conclude that
the conductance dips correspond to the inter-subband
transitions that occur with probability one.
On the contrary, for the magnetic fields (III) and (V) 
the probability of the inter-subband transition is less than one.
Although the main conductance dip for $B_{ext}=B_h$ [point (IV)] can be
easily interpreted in terms of the ordinary Landau-Zener
transition,\cite{Landau, Zener, Saarikoski2014,Betthausen2012} the
occurrence of the dips for $B_{ext}\ne B_h$ cannot be explained by the
existing theory.

\begin{figure}[ht]
\begin{center}
\includegraphics[scale=0.25, angle=0]{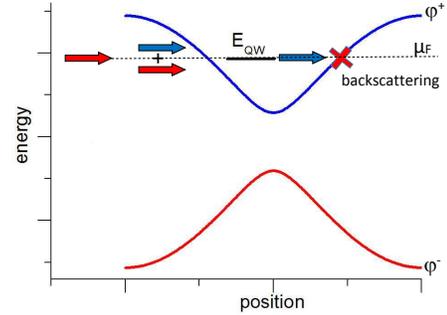}
\caption{(Color online) Schematic illustration of the resonant
Landau-Zener transition leading to the backscattering and 
additional conductance dips for $B_{ext}\ne B_h$. Colors of the arrows
correspond to the electron eigenstates: red corresponds to the
state with the spin parallel to the magnetic field and blue corresponds
to the state with the spin antiparallel to the magnetic field.}
\label{fig6}
\end{center}
\end{figure}

\begin{figure}[ht]
\begin{center}
\includegraphics[scale=0.25, angle=0]{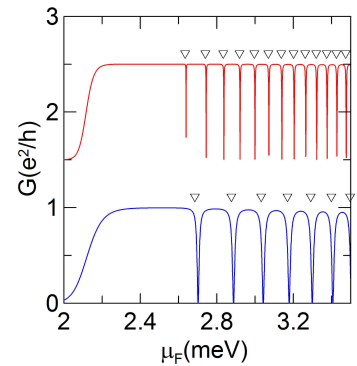}
\caption{(Color online) Conductance $G$ as a function of Fermi
energy $\mu _F$ calculated  for $B_{ext} = 70$~mT and for $a=1$ $\mu$m
(blue curve) and $a=2$ $\mu$m (red curve, for a better visibility
shifted upwards by a unit of conductance). The triangles correspond to
the energy levels of the quasi-bound states  in the effective quantum
well created in subband $\varphi ^{+}$.}
\label{fig7}
\end{center}
\end{figure}

The physics behind this effect can be explained as follows.
If the period of the spatial magnetic field modulation is short,
the adiabaticity condition for the spin transport through the
nanostructure is violated. In this non-adiabatic regime, the state of
the electron, initially injected into the lowest-energy subband $\varphi
^{-}$ is a linear combination of the eigenstates with the spin
orientation parallel and antiparallel to the magnetic field. This causes
that the probability of the inter-subband transition is non-zero. If
Fermi energy $\mu _F$ (the energy of the injected electrons) exceeds
$E^{+}_{min}$, the electrons flowing through the nanostructure
experience the effective quantum well created in the spatially varying
subband $\varphi ^{+}$.
In this quantum well, the quasi-bound electron states are created.
If the Fermi energy becomes equal to the energy of one of the
quasi-bound states formed in subband $\varphi ^{+}$, the probability
of inter-subband transition approaches one.  This effect, being
analogous to the resonant tunneling, can be called the resonant
Landau-Zener transition. As a result of these resonant transitions all
electrons injected into the nanowire are transmitted to the subband
$\varphi^{+}$.  However, since the energy of the subband $\varphi ^{+}$
in the right contact is higher than the Fermi energy, the electrons are
backscattered, which gives raise to the conductance dip.  This mechanism
is schematically illustrated in
Fig.~\ref{fig6}.
In order to show that the resonant Landau-Zener transition
leads to the conductance dip for $B_{ext} \ne B_h$,
we display in Fig.~\ref{fig7} the conductance as a function of the Fermi
energy. The  triangles mark the numerically calculated energies of the
quasi-bound states in the effective quantum well created in the subband
$\varphi ^{+}$. We see that these energy levels agree very
well with the positions of the conductance dips. The results of
the extended calculations of the quasi-bound state energy levels as
functions of the magnetic field $B_{ext}$ and the Fermi energy $\mu _F$
are presented in Fig.~\ref{fig2}(b) by white dashed curves.
The good agreement between the positions of the conductance dips and
the energies of the quasi-bound states allows to conclude that the
conduction dips for $B_{ext} \ne B_{h}$ result from the resonant
Landau-Zener transition between the corresponding subbands.
We note that the necessary condition for this effect is the
non-adiabaticity of the transport (in our case resulting from the
sufficiently short period $a$ of the magnetic field spatial modulation),
because only in this regime the state of the electron is the linear
combination of the two eigenstates with the opposite spins. In the
adiabatic regime, the electron injected into the lowest-energy subband
$\varphi ^{-}$ remains in this state when flowing through the entire
nanostructure.  Therefore, the probability of the Landau-Zener
transitions is zero. This argument explains why the  effect presented in
this paper disappears for the long period of the magnetic field
modulation [see Fig.~\ref{fig2}(c)], for which the transport can be
treated as adiabatic.

In summary, we have demonstrated that the current flowing through the
waveguide (nanowire) made of the magnetic semiconductor in the
helical magnetic field exhibits the additional conductance dips for
$B_{ext} \ne B_{h}$. This effect has been explained as resulting form
the resonant Landau-Zener transitions between the spin-splitted subbands
that leads to the  spin backscattering. We believe that our findings can
be important for the spin transistor design based on helical magnetic
field. 
Until now the transport regime in the new design is restricted to
the adiabatic regime. We have shown the in the non-abiadatic regime
(for the sufficiently small distance between the ferromagnetic stripes)
the spin transistor action can be also induced by the resonant
Landau-Zener transitions.

This work has been supported by the National Science Centre, Poland,
under grant DEC-2011/03/B/ST3/00240.

%

\end{document}